\documentclass[fleqn,usenatbib]{mnras}
\usepackage{newtxtext,newtxmath}
\usepackage[T1]{fontenc}
\DeclareRobustCommand{\VAN}[3]{#2}
\let\VANthebibliography\thebibliography
\def\thebibliography{\DeclareRobustCommand{\VAN}[3]{##3}\VANthebibliography}

\newcommand{\abs}[1]{\lvert #1 \rvert}

\usepackage{ae,aecompl}
\usepackage{float}
\usepackage{graphicx}   

\usepackage{color}

\title[Photometric $\mathrm{[Fe/H]}$ of Oo\,I/II RRab stars]{On the photometric metallicity of Oosterhoff-type I and II RRab stars}

\author[J. Jurcsik et al.]{
J. Jurcsik,$^{1}$\thanks{E-mail: jurcsik.johanna@gmail.com}
G. Hajdu$^{2}$, \'A. Juh\'asz$^{1}$\\
$^{1}$Konkoly Observatory, H-1121 Budapest, Konkoly Thege Mikl\'os \'ut 15-17., Hungary\\
$^{2}$Nicolaus Copernicus Astronomical Center, Polish Academy of Sciences, Bartycka 18, 00-716, Warsaw, Poland\\
}
\begin{document}
\date{Accepted 2021 May 11. Received 2021 May 5; in original form 2021 Marc 22}
\pagerange{\pageref{firstpage}--\pageref{lastpage}} \pubyear{2021}
\maketitle

\label{firstpage}

\begin{abstract}
We are revising the consistency of photometric metallicity formulae widely used for fundamental-mode RR Lyrae (RRab) variables, based on their $V$- and $I$-band light curves, published by Jurcsik and Kov\'acs (1996) and Smolec (2005), respectively. 293 RRab variables belonging to 10 globular clusters, all simultaneously containing stars of both Oosterhoff types, are selected for this purpose. We find that on average, the $V$-band formula results in higher estimated metallicities by about 0.05\,dex than the $I$-band formula. Moreover,  we detect a dependency on the Oosterhoff class of the variables for both formulae, as well. Using the $V$-band formula, Oo\,I stars are $0.05-0.10$ dex more metal rich than Oo\,II stars of the same cluster. Although with less significance, but the $I$-band results indicate a reversed trend. Therefore, we surmise that the average difference we have found between the $V$- and $I$-band formulae is the consequence of the total sample being dominated by Oo I variables.

\end{abstract}

\section{Introduction}\label{intro}

Stellar metallicity is one of the most crucial parameters in astronomical, astrophysical and cosmological studies, as well.
Direct spectroscopic metallicity determinations are often prohibitively costly for large samples of very far, faint objects. Therefore, it is not surprising that several indirect methods using photometric information have been developed to estimate the metallicities of different type stars. 

Concerning RR Lyrae (RRL) stars, the most widely used technique is to calculate the [Fe/H] from simple parameters obtained from the Fourier solution of the light curve. The linear formulae published in \cite{jk96} and in  \cite{s05}, which use $V$ and $I$-band light-curve parameters of fundamental-mode RR Lyrae stars are estimated to give  [Fe/H] values with $0.13-0.18$ dex accuracy. \cite{nemec} derived a quadratic formula for the photometric band of the Kepler satellite, and \cite{haj18} extended the method to the near-infrared $K_\mathrm{s}$ photometric band.\footnote{At the time of submitting the paper new photometric metallicity formulae for optical and infrared bands were published by \cite{mul21}.} \cite{mor07} defined a similar relation to estimate the [Fe/H] of overtone RRc variables.

Because of their ease of use and the access to large amount of light-curve data provided the modern small/medium-size robotic telescopes, the application of these formulae are very popular \citep[{see e.g.}][]{tor15,skow16,fer17,jac20,piet20}. However, there are some problems related to the limitations, compatibility and inherent accuracy of these formulae. For example, the photometric [Fe/H]  overestimates the true cluster metallicity for the most metal-poor globular clusters (GCs) as it was already noticed in \cite{jk96} \citep[see also  e.g. NGC 5053,][]{nem04}, which might be connected with some systematic bias between the results for Oosterhoff-type I and type II stars (Oo\,I/Oo\,II), as these are typical Oo\,II clusters.

The Oosterhoff dichotomy, detected in globular cluster RR Lyrae stars \citep{oo39,arp55}, has still not been expained in full details. For an overview of the subject, see \cite{cat09}. 
The Oo classification of Galactic GCs is based on some properties of their RRL populations:  e.g. the mean period of the RRab stars, their location on the Bailey diagram ($period-amplitude$ plot) and the number ratio of RRab to RRc stars. There is a so called period shift between similar amplitude Oo\,I and Oo\,II stars  and vice versa, the amplitude of an Oo\,II star is larger than the amplitude of a similar period  Oo\,I variable. Therefore, the compatibility of the two-parameter $V$-band formula, which does not involve amplitude information, and the three-parameter $I$-band formula, which also includes the $A_2$ Fourier amplitude with a relatively large, 7.9, coefficient accounting for about 1.2 dex changes in the results over the possible range of the $I$-band $A_2$ amplitudes, might be  suspicious, too. 

We note here, however, that the addition of an amplitude term to the metallicity formula in the $V$-band did not improve the fit at all \citep{jk96}. The s.dev. of the two- and three-parameter $V$-band formulae are 0.134 and 0.132 dex. This is in contrast with the $I$-band results, where the difference between the accuracy of the two- and three-parameter formulae is statistically significant (0.18 / 0.14 dex), and the three-parameter formula gives results of similar accuracy as the two-parameter $V$-band formula.   Moreover, it was shown in \cite{haj18} that ``the metal abundance formula described by eq.~2 of \cite{s05} suffers from a systematic bias as a function of amplitude/period", and that   eq.~3  ``gives much more consistent abundance estimates as a function of the amplitude." As the amplitude-dependent residual using eq.~2 of \cite{s05} was found to be larger than 0.3 dex \citep[see fig. 10 in][]{haj18} we do not consider this formula to yield reliable results and restrict the investigation only to eq. 3 of \cite{s05}. This decision is also justified in Sect.~\ref{OoI-II}.

Using OGLE IV photometry \citep{sos17}, a systematic difference between the results for the Oo\,I and Oo\,II type samples in the Galactic bulge was detected by \cite{prud19}. The median [Fe/H] value for the Oo\,II stars proved to be 0.1 dex more metal poor than for the Oo\,I stars. 
However, \cite{haj18} noticed a $\sim0.1$\,dex amplitude-dependent residual in the [Fe/H] values derived for Oo\,I variables with eq.~3 of \cite{s05} in the Galactic bulge RR Lyrae sample. The correction for this effect results in somewhat lower metallicities for Oo\,I RR Lyrae. Nevertheless, these type of corrections hasn't been widely adopted for the derivation of photometric metallicity estimates, we are focusing on verifying the behaviour of the  original eq.~3 \cite{s05} formula.

Somewhat surprisingly, neither the $V$- nor the $I$-band formulae have never been scrutinised for systematic biases between the photometric [Fe/H] of Oo\,I and Oo\,II stars. In order to alleviate possible inconsistencies of the method, we decided to perform a systematic check using mono-metallic GC data.

The primary aims of the present study are: 

1) To verify the compatibility of the $V$- and $I$-band results: do the $V$- and $I$-band formulae indeed provide identical results, and does this extend to both Oo\,I and Oo\,II variables?

2) To revise the compatibility of results obtained for Oo\,I- and Oo\,II-type stars: are the photometric [Fe/H]  values of Oo\,I- and Oo\,II-type stars of mono-metallic systems like GCs consistent? This is checked for the $V$- and $I$-band data separately.

\section{The metallicity scales}\label{scale}

\begin{figure}
\centering
\includegraphics[width=8.5cm]{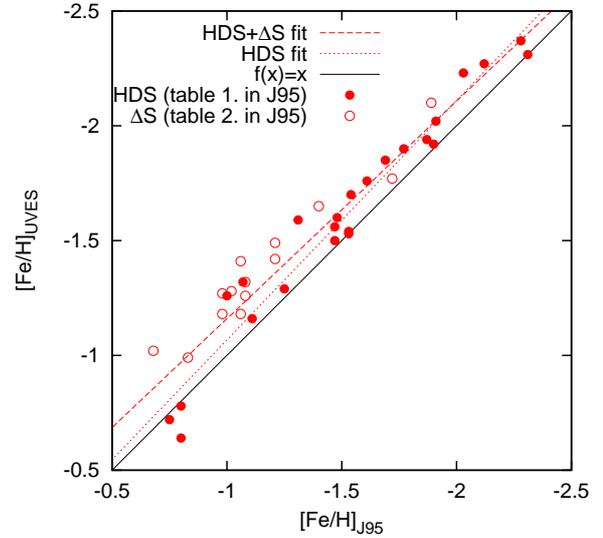}
\caption{The [Fe/H] values (HDS, $\Delta$S data), which were used to define the J95 metallicity scale of Galactic GCs, and their  [Fe/H]$_{\mathrm{UVES}}$ values \citep[][2010 edition]{harris} are compared. Open and filled symbols indicate cluster metallicities according to the data given in table 1 (HDS data) and table 2 ($\Delta$S measurements) in J95. Clusters with larger than 0.5 dex differences between the UVES and JK95 metallicities are omitted. The linear fits to all data and to the HDS values alone are indicated by dashed and dotted red lines, respectively.  The black line corresponds to the equality relation. }
\label{scale.fig} 
\end{figure}

In principle, both photometric metallicity values calculated using the $V$- and the $I$-band light-curve parameters using the formulae published by  \cite{jk96} and \cite{s05}, respectively, yield results on the metallicity scale defined in \citet[][hereafter J95]{phot}. Concerning GCs, this metallicity scale is based on high-dispersion spectroscopic (HDS) data and on the $\Delta$S measurements  available at the mid-1990's.

The currently most commonly employed  metallicity scale  (${{\mathrm{[Fe/H]}}_{\mathrm{UVES}}}$) of GCs has been defined in \citet{carretta}. This metallicity scale is adopted for the compilation of GC data in  the electronic edition (2010) of the \cite{harris} catalogue. 
Correlating the ${{\mathrm{[Fe/H]}}_{\mathrm{UVES}}}$  metallicities \citep{carretta} with the  ${{\mathrm{[Fe/H]}}_{\mathrm{HDS}}}$ metallicities of 24 clusters, which have been utilised, partly,  to define the J95 scale, \citet{haj18} derived a linear transformation between the two metallicity scales. However, if GCs with  $\Delta$S metallicities (table 2 in J95) are also considered, a somewhat different relation between the UVES and the J95 scales is derived.

Fig.~\ref{scale.fig} compares the [Fe/H]$_{\mathrm{J95}}$ and the [Fe/H]$_{\mathrm{UVES}}$ values of Galactic GCs used to set the J95 metallicity scale.  The linear fits for clusters listed exclusively in table 1 \citep[eq.~6 in][]{haj18} and for data of both table 1 and table 2 of J95 are shown by dotted and dashed lines, respectively. Similarly as in \citet{haj18}, GCs with larger than 0.5 dex differences between the two metallicities are omitted. There is one such cluster in the HDS sample (NGC\,5927) and there are three GCs with $\Delta$S metallicities (NGC 6284, NGC\,6712 and NGC\,6723).   Most probably the HDS and  $\Delta$S metallicities of these clusters used by J95 were not accurate enough.

The photometric metallicities of the GCs involved in the present study transformed to the UVES scale according to the linear relation defined using both the   ${{\mathrm{[Fe/H]}}_{\mathrm{HDS}}}$ and $\Delta$S data listed in J95 (red dashed line in Fig~\ref{scale.fig}) are in a better agreement with their ${\mathrm{[Fe/H]}}_{\mathrm{UVES}}$-scale values than if using equation 6 of \cite{haj18}. 
The mean differences between the [Fe/H]$_{\mathrm{UVES}}$ and the median values of the cluster photometric metallicities using equation 6 of \cite{haj18} are $0.133\pm0.173$ and  $0.091\pm 0.189$ in the $V$ and $I$ bands, respectively, while these values are $0.087\pm.185$ and $0.049\pm.201$ when using the transformation that relies on the $\Delta$S measurements, as well.

Therefore, we accept the following transformation between the UVES and J95 scales (solid line in Fig.~\ref{scale.fig}):
\begin{equation}\label{scale.eq}
{{\mathrm{[Fe/H]}}_{\mathrm{UVES}}}= 0.948 {{\mathrm{[Fe/H]}}_{\mathrm{J95}}} - 0.211.
\end{equation}
The corresponding modified forms of the  \cite{jk96} and \cite{s05} formulae, which already yield  [Fe/H] values on the UVES scale  are the following:
\begin{equation}\label{fe/h.v.eq}
{\mathrm{[Fe/H]}}_V=-4.987-5.114P+1.275\varphi_{31} 
\end{equation}
\vskip -15pt
\begin{equation}\label{fe/h.i.eq}
{\mathrm{[Fe/H]}}_I=-6.018-4.261P+1.120\varphi_{31}+7.466A_2 
\end{equation}
We note here, however, that these modifications of the photometric formulae may not be valid for metal-rich field stars, but the compatibility with these variables is out of the scope of this paper.

The photometric metallicity calculated according to Eqs.~\ref{fe/h.v.eq} and \ref{fe/h.i.eq}, which is scaled to match the ${\mathrm{[Fe/H]}}_{\mathrm{UVES}}$ scale of GCs, is denoted as {$\mathrm{[Fe/H]}_{\mathrm{phot}}^*$} hereafter. These values are about 0.1 dex more metal-poor as the original [Fe/H]$_{\mathrm{J95}}$ values for the most metal-poor clusters ([Fe/H]$ < -2$), and the difference is  $-0.16$ for clusters with [Fe/H]$\sim -1$. 
The application of this transformation, however,  does not solve the problem of the overestimation of the photometric [Fe/H] for the most metal-poor GC,  as the results are still not metal-poor  enough for these clusters.

\section{The sample}

\begin{table*} 
\begin{center} 
\caption{Mean and median photometric metallicities of GCs calculated using the $V$- and $I$-band light-curve parameters of RRab stars.\label{ered1.tab}} 
\begin{tabular}{l@{\hspace{0pt}}c@{\hspace{6pt}}lr@{\hspace{6pt}}rr@{\hspace{6pt}}rcr@{\hspace{6pt}}rr@{\hspace{6pt}}rc@{\hspace{6pt}}r@{\hspace{6pt}}l}
\hline
Name& \multicolumn{1}{c}{Oo-type$^a$} &{${\mathrm{[Fe/H]}}^b$}&\multicolumn{2}{c}{${\mathrm{[Fe/H]}}_{V}$ }&\multicolumn{2}{c}{${{\mathrm{[Fe/H]}_{V}^*}}$}&S.dev.&\multicolumn{2}{c}{${\mathrm{[Fe/H]}}_{I}$}&\multicolumn{2}{c}{${{\mathrm{[Fe/H]}_{I}^*}}$}&S.dev.&N&Ref.$^c$\\
&&mean&med.&mean&med.&&mean&med.&mean&med.&&&\\
\hline
M68	       &II &-2.23$^{\dagger}$	&-1.688	&-1.712	&-1.811	&-1.834 &0.139	&-1.741	&-1.682	&-1.861	&-1.805&0.147 &	7 &1,2\\
M53	       &II &-2.10	        &-1.576	&-1.602	&-1.705	&-1.730 &0.095	&-1.603	&-1.583	&-1.730	&-1.711&0.128 &	24&3,4\\
NGC\,3201      &I  &-1.59$^{\dagger}$	&-1.264	&-1.263	&-1.409	&-1.408 &0.121	&-1.362	&-1.367	&-1.502	&-1.507&0.132 &	29&5,6\\
IC4499         &I  &-1.53	        &-1.477	&-1.472	&-1.611	&-1.606 &0.158	&-1.517	&-1.496	&-1.649	&-1.629&0.195 &	32&\\
M3	       &I  &-1.50          	&-1.351	&-1.364	&-1.492	&-1.505 &0.105	&-1.391	&-1.382	&-1.529	&-1.522&0.099 &	62&8\\
NGC\,6934      &I  &-1.47               &-1.244	&-1.273	&-1.390	&-1.417 &0.210	&-1.354	&-1.411	&-1.495	&-1.548&0.215 &	27&9,10\\
NGC\,6229      &I  &-1.47$^{\dagger}$	&-1.132	&-1.163	&-1.285	&-1.313 &0.131	&-1.242	&-1.247	&-1.388	&-1.393&0.152 &	20&11\\
M5	       &I  &-1.29	        &-1.160	&-1.169	&-1.311	&-1.320 &0.136	&-1.214	&-1.221	&-1.362	&-1.368&0.102 &	48&12,13,14\\
M14	 &\,\,I$^{\ddagger}$ &-1.28	&-1.191	&-1.142	&-1.340	&-1.294 &0.204	&-1.286	&-1.255	&-1.430	&-1.401&0.222 &	32&15\\
NGC\,1851      &I  &-1.18$^{\dagger}$	&-1.178	&-1.191	&-1.328	&-1.340 &0.130	&-1.122	&-1.112	&-1.275	&-1.265&0.112 &	12&16\\
\hline
\multicolumn{15}{l}{$^a$ Oosterhoff type given in \cite{cast03}.}\\
\multicolumn{15}{l}{$^b$ ${{\mathrm{[Fe/H]}}}$ data on the UVES scale \citep{carretta} are taken from the 2010 edition of the \cite{harris} catalogue.}\\
\multicolumn{15}{l}{$^c$ References to the photometric data: 1) \cite{kai15}; 2) \cite{wal94}; 3) \cite{fer12}; }\\
\multicolumn{15}{l}{\,\,\, 4) \cite{dek09} only for V14 and V28; 5) \cite{fer14}; 6) \cite{lay03}; 7) \cite{wal96};}\\
\multicolumn{15}{l}{\,\,\, 8) \cite{jur17}; 9) \cite{ye18}; 10) \cite{kal01}; 11) \cite{fer15}; 12) \cite{kal00}; }\\
\multicolumn{15}{l}{\,\,\,   13) \cite{fer16}; 14) \cite{reid} data is also used for V36; 15) \cite{cont18}; 16) \cite{wal98}.}\\
\multicolumn{15}{l}{$^*$ The superscript $^*$ denotes photometric metallicities transformed to the UVES scale according to Eq. 1.} \\
\multicolumn{15}{l}{$^{\dagger}$ Larger than 0.1 dex different [Fe/H] values of these clusters were also published in the literature: $-2.42$ \citep[M68,][]{sch15};}\\ 
\multicolumn{15}{l}{\,\, $-1.47$  \citep[NGC\,3201,][]{mag18}; $-1.13$  \citep[NGC\,6229,][]{john17}; $-1.33$ \citep[NGC\,1851,][]{mar14}.}\\
\multicolumn{15}{l}{$^\ddagger$  \cite{cont18} classified M14 as an Oo intermediate cluster.}
\end{tabular}
\end{center}
\end{table*}

Time-series data of  RRL variables  in both the $V$- and the  $I$-bands of Galactic GCs  were obtained from the literature. Each GC with relatively large known population of RRL stars was checked. Systematic biases in the photometric metallicities of RRL variables belonging to the two Oo classes can be detected most reliably  using  samples of stars with homogeneous metallicities. Therefore $\omega$ Cen and other clusters with detected metallicity spread like e.g. M2 \citep{yong14} and M19 \citep{john15} has been excluded from the study.

 The Bailey diagram  of the selected clusters was inspected in order to restrict the investigation to clusters that contain RRab stars of both Oo types. Although the Galactic GCs are categorised to belong to one of the Oo types on the base of some properties of their RRL populations, most of the clusters contain some RRLs belonging to the other Oo class, too \citep[see e.g. M3,][]{j03}. The division of GCs into Oo\,I and Oo\,II types disappear in extragalactic GCs. Their properties indicate Oo-intermediate type classification \citep{bon94,cat09}. M14, which might have extragalactic origin,  is also an Oo$_{\mathrm {int}}$ cluster according to \cite{cont18}.

Variables with light curves suitable for the determination of photometric metallicities in both bands are selected. This makes an unbiased comparison of the  $V$- and the $I$-band results possible.  Long-period variables with sinusoidal-shape light curves, like V203 in M3, are not used because their $\varPhi_{31}$ parameter is too uncertain to determine reliable photometric metallicities. Variables with significantly different $V$- and $I$-band photometric metallicities  ($\abs{{\mathrm{[Fe/H]}}_{V}-{\mathrm{[Fe/H]}}_{I}}>0.5$) and variables located significantly below the Oo\,I ridge on the Bailey diagram \citep[e.g. V14,V35,V37,V40,V46,V49 and V53 in NGC\,6229,][]{fer15} are removed from the sample. We surmise that these stars have defective photometries,  most probably because crowding biased the light curves differently in the different bands.

When data from different sources of a given star are used, the zero points of the data sets are matched if necessary. 
Zero-point offsets are also detected between different-season observations of the same data set in some cases \citep[e.g. in M53,][]{fer12}. These are also corrected.
The correction means, simply, that the mean magnitudes of the different parts of the data are matched, without any consider on the direction of the offset. However, as the mean magnitudes of the variables are not involved in any context of the present work, the correctness of the magnitude zero points is indifferent from our points of view.

 Extended, multi-colour, good quality time-series data are still not available for many GCs. Therefore, in order to collect a  large enough data set to draw any conclusion, we did not restrict the sample to the best quality, non-modulated variables with full phase coverage.

Although the sparseness and/or shortness of the data sets do not make it possible to identify the Blazhko stars in some of the clusters,  stars with large modulations, if detected, are omitted from the investigation as their photometric metallicity may not be reliable.  However, in order to increase the sample size, RRab stars are included if the amplitude of the modulation, both in amplitude and in phase, is small.  These stars can be identified without doubt only in the M3 data. In general, the light curves are too scattered to distinguish small-amplitude modulation from photometric errors, or the different band or different source of observations are contradictory regarding the modulation. With the exception of M3, the number of stars with evident small-amplitude modulation included in the sample is very few. The $V$-band photometric metallicities of the known Blazhko and non-Blazhko M3 variables involved in the analysis do not show any systematic differences. The mean values are $-1.35$ s.dev. 0.11 (10 stars) , and $-1.34$ s.dev. 0.12 (52 stars), respectively.

Outlying/erroneous data points are removed from the time-series data.
The light curve of some stars is too sparse to obtain a reliable, high enough order Fourier fit. In order to stabilise the Fourier parameters, the gaps in the light curves of these stars are filled by a few ($2-6$) interpolated artificial data points.  Different solutions were checked, and if the derived photometric metallicities were divergent, the star was omitted from the sample. In some, very limited cases, when the minimum-maximum phases were not completely covered, the shape of the light curve in other bands (if available) and of similar period and amplitude variables in the cluster were used for guidance.

 The Fourier parameters, which are involved in calculating the photometric metallicity values, are determined star by star, using appropriate-order harmonic fit to the data.  In order to fit the steepest and most curving parts of the light curves correctly, the highest  possible order was chosen, which still yielded a smooth and not winkled solution.

The final sample of stars suitable for the analysis consists of 293 RRab stars belonging to 10 Galactic GCs. 
The cluster names, their Oo classification, the ${{\mathrm{[Fe/H]}}_{\mathrm{UVES}}}$ values  \citep[2010 edition of][]{harris}, the mean and the median values of the $V$- and $I$-band photometric metallicities according to the  \cite{jk96} and \cite{s05} formulae, and these values  transformed to the UVES metallicity scale as given in  Eqs.~\ref{fe/h.v.eq}, and ~\ref{fe/h.i.eq}, the standard deviation of the photometric metallicities in the $V$ and $I$ bands, the number of the RRL light curves used for the analysis, and the references of the photometric data are given in the columns of Table~\ref{ered1.tab}.

The photometric data are in the Johnson-Cousins system. The instrumental magnitudes  of all  utilized observations were tied to the standard system using photometric sequences published by \cite{landolt83,landolt92} or \cite{stet00}.

To indicate the uncertainty of the catalogue metallicity values of the GCs, some recent [Fe/H] measurements, differing by more than 0.1 dex from the catalogue value, are also given in Table~\ref{ered1.tab}. 

\section{Comparison of the  $V$- and $I$-band photometric metallicities}\label{mean1}

\begin{figure*}
\centering
\includegraphics[width=17.5cm]{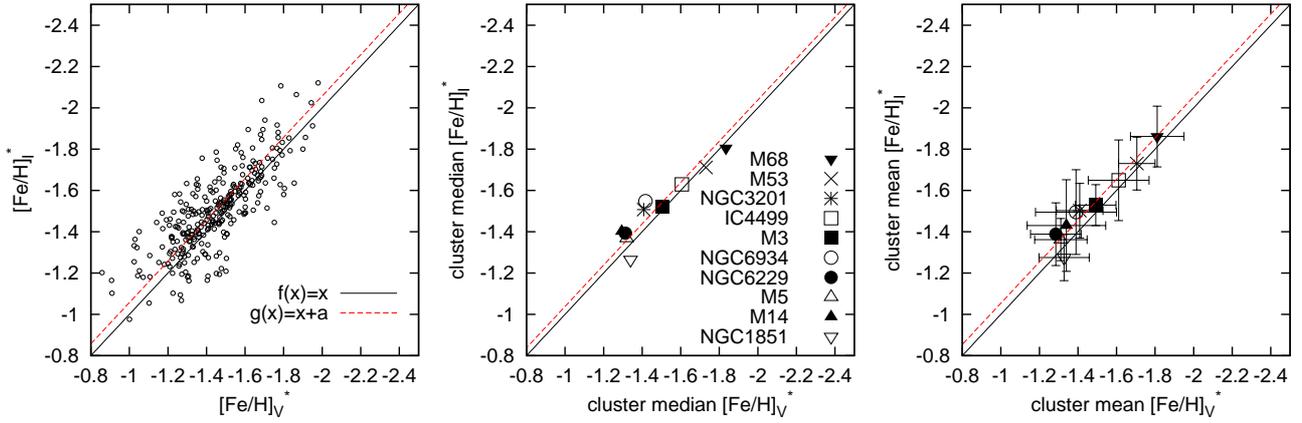}
\caption{The comparison  of the $V$- and $I$-band photometric metallicities of the 293 stars is shown in the left-side panel.  The middle and right-hand panels show the results for the cluster-median and -mean photometric metallicity values.   The error bars indicate the s.dev. of the derived photometric metallicies of the clusters in the right-hand panel. The solid black lines show the equality relation and the fits with constant shift between the data are shown by dashed red lines in the plots. The photometric metallicities are on the UVES scale. }
\label{meanvas1.fig} 
\end{figure*}
\begin{figure*}
\centering
\includegraphics[width=17.5cm]{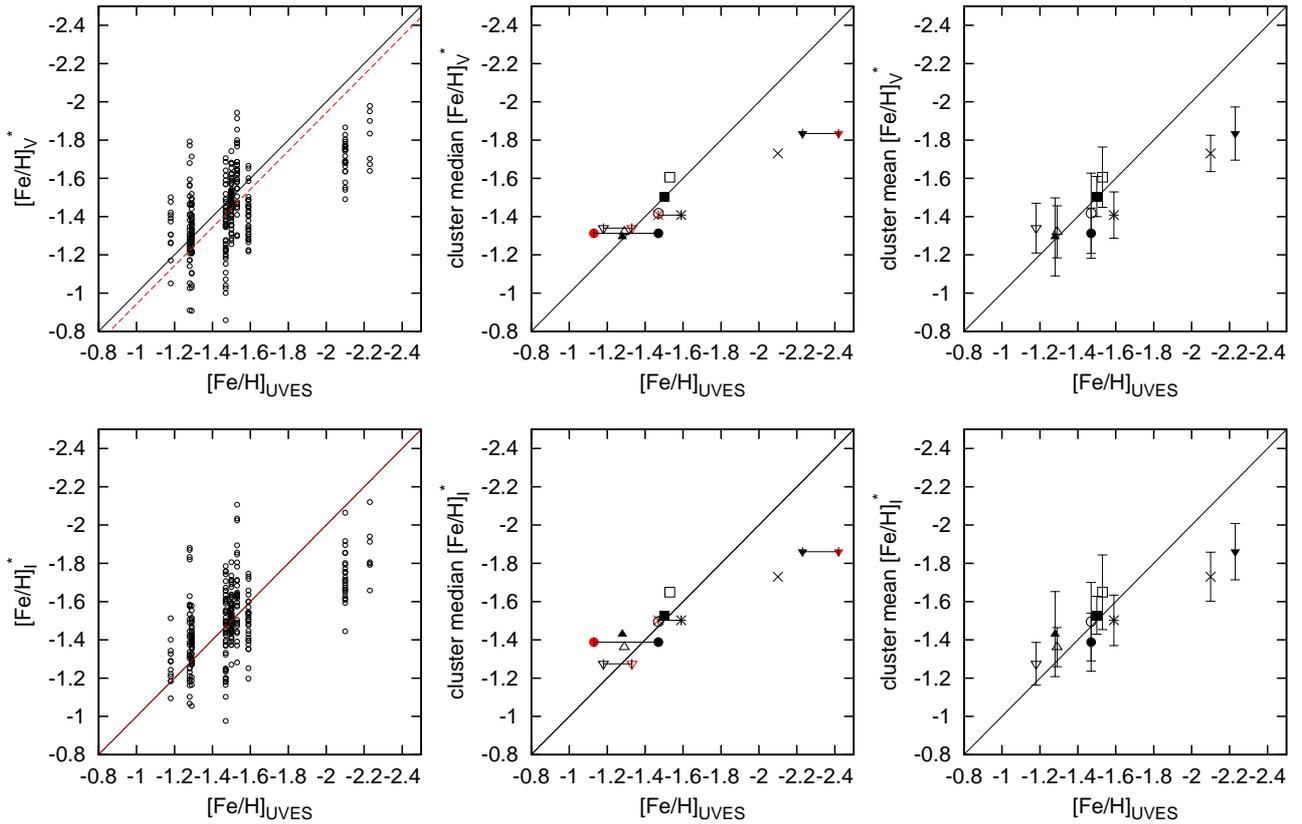}
\caption{The  $V$-band (top panels) and $I$-band (bottom panels) photometric metallicities are compared with the ${{\mathrm{[Fe/H]}}_{\mathrm{UVES}}}$ values of the clusters given in the 2010 edition of the  \citet{harris} catalogue. Results for all the stars and the cluster median and mean values are plotted in the left-side, middle and right-side panels, respectively.
 Some recent metallicity measurements  of the clusters listed in the footnote of Table~\ref{ered1.tab} are also plotted by red symbols connected with horizontal lines to the catalogue values in the middle panels in order to indicate the uncertainties of the catalogue values.
 The line types and symbols shown are the same as in Fig.~\ref{meanvas1.fig}. }
\label{meanvas2.fig} 
\end{figure*}

\begin{figure*}
\centering
\includegraphics[width=18cm]{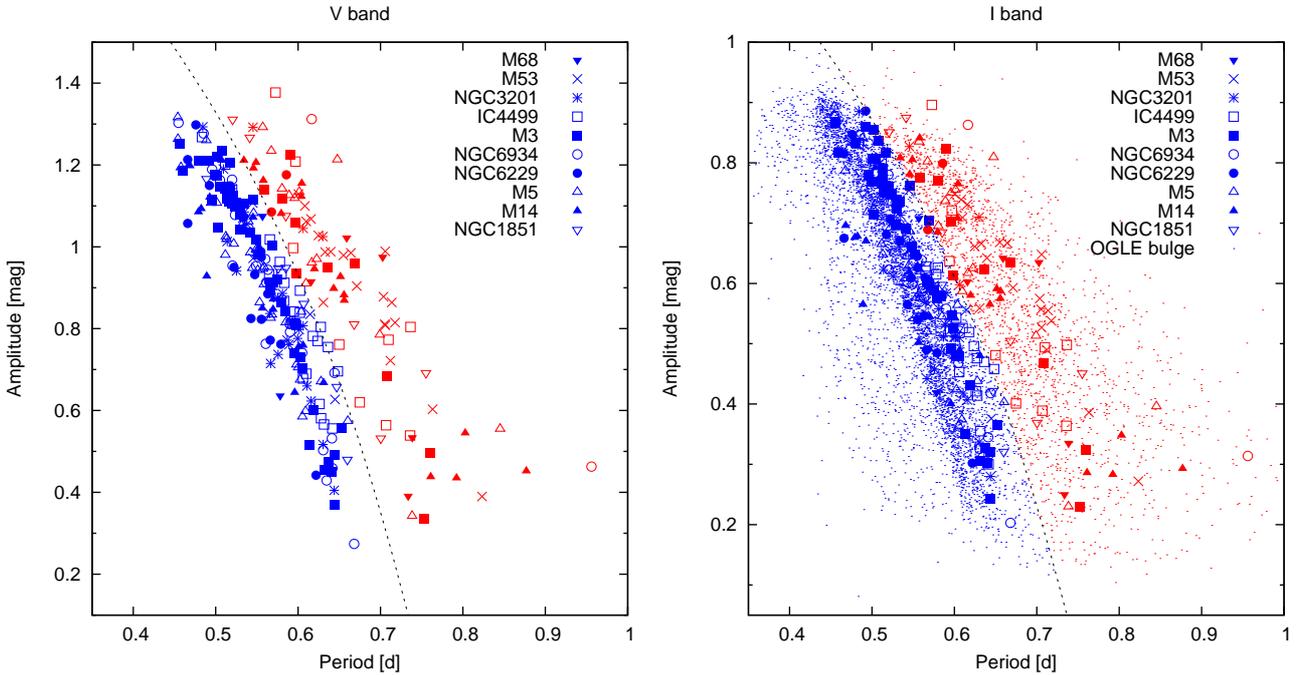}
\caption{Bailey diagrams of the sample stars in the $V$ (left-side panel) and the $I$ (right-side panel) bands.  Stars in different clusters are denoted by different symbols. Oo-type I and II stars are shown in blue and red, respectively. The OGLE-IV Galactic bulge Oo\,I and Oo\,II stars \protect\citep{prud19} are also displayed by small dots in the right-side panel. The borderlines between the two Oo types determined from the OGLE bulge data  (Eq. \ref{ifit.eq}) and transformed to the $V$ band (Eq. \ref{vfit.eq}) are shown by dotted lines in the panels.}
\label{Bailey.fig} 
\end{figure*}

Fig.~\ref{meanvas1.fig}  and Fig.~\ref{meanvas2.fig} document the results for the complete sample of the cluster stars (left-hand panels) and for the cluster median (middle panels) and mean (right-hand panels) photometric [Fe/H]  values calculated according to Eqs.~\ref{fe/h.v.eq}, and ~\ref{fe/h.i.eq}. The error bars in the right-hand  panels of the figures correspond to the standard deviation of the photometric metallicities of the cluster RRab stars.  Black solid lines  indicate equality of the data in each panels of the figures.

The compatibility of the  $V$- and $I$-band photometric [Fe/H] results is shown in Fig.~\ref{meanvas1.fig}. T Fitting the data supposing a constant offset between the metallicity values derived from the  $V$- and $I$-band observations on each panel in Fig.~\ref{meanvas1.fig}  (red dashed lines), the result indicates that the $I$-band formula yields a slightly lower metallicity value than the $V$-band formula. According to the fits to the sample of the 293 stars, the cluster-median and -mean values, the differences are $0.057\pm0.008$, $0.038\pm0.021$ and $0.054\pm0.015$, respectively. Based on these values, we conclude that there is a systematic, $\approx0.055$ dex difference between the results obtained utilising the $V$-  and $I$-band  metallicity formulae, being the results in the  $I$ band of slightly lower metallicity than in the $V$ band.

Fig.~\ref{meanvas2.fig} shows the $V$- and $I$-band photometric metallicities as a function of the ${{\mathrm{[Fe/H]}}_{\mathrm{UVES}}}$ values of the clusters given in the 2010 edition of the  \citet{harris} catalogue. 
Some more recent metallicity measurements of the clusters, given in the footnotes of Table~\ref{ered1.tab}, which differ by more than 0.1 dex from the catalogue values, are also shown  in the middle panels (red symbols), in order to highlight the uncertainties of the literature cluster metallicities.

The photometric metallicities obtained in the $V$- and $I$-bands are  in acceptable agreement with the cluster metallicities when taking into account  both the possible uncertainties of the catalogue values, and the standard deviation of the photometric [Fe/H] values, with the exceptions of the two most metal deficient clusters of the sample, M53 and M68. However, that  the photometric formulae overestimate the metallicity by $0.3-0.5$ dex for the most metal-poor clusters is not a new result,  as it was already mention in Sect.~\ref{intro}.

Calculating the mean difference between the catalogue values and the photometric metallicities (red dashed lines in the left-side panels in Fig.~\ref{meanvas2.fig}) no difference at all ($0.000\pm 0.012$) in the $I$-band and a  $0.057\pm0.012$ shift in the $V$ band are derived. This is exactly  the same as the difference between the $V$- and $I$-band results as documented in Fig.~\ref{meanvas1.fig}. 

We conclude that, to match the photometric metallicities to the${{\mathrm{[Fe/H]}}_{\mathrm{UVES}}}$ values, the application of an additional $\sim0.05$~dex correction of the $V$-band results (obtained using Eq. \ref{fe/h.v.eq}) would be necessary.

\section{Separation of the two Oosterhoff types}

\begin{table*} 
\begin{center} 
\caption{Mean and median values of the photometric $V$- and $I$-band metallicities of Oo\,I and Oo\,II variables in 10 GCs.\label{ered2.tab}} 
\begin{tabular}{lcrrrr@{\hspace{20pt}}rrrrrr}
\hline
Name&& \multicolumn{4}{c}{Oosterhoff I}&\multicolumn{4}{c}{Oosterhoff II} &N$_{\mathrm{Oo\,I}}$ &N$_{\mathrm{Oo\,II}}$ \\
&{${\mathrm{[Fe/H]}}_{\mathrm{UVES}}$}&\multicolumn{2}{c}{${\mathrm{[Fe/H]}_{V}^{*}}$}& \multicolumn{2}{c}{${\mathrm{[Fe/H]}_{I}^{*}}$}&\multicolumn{2}{c}{${\mathrm{[Fe/H]}_{V}^{*}}$}&\multicolumn{2}{c}{${\mathrm{[Fe/H]}_{I}^{*}}$} &&\\
&&mean&med.&mean&med.&mean&med.&mean&med.&&\\
\hline
M68	        &-2.23	&-1.656	&-1.656	&-1.732	&-1.732	&-1.873	&-1.900	&-1.913	&-1.912	&2&5 \\
M53	        &-2.10	&-1.740	&-1.730	&-1.893	&-1.906	&-1.698	&-1.730	&-1.698	&-1.697	&4&20\\
NGC\,3201	&-1.59	&-1.401	&-1.390	&-1.516	&-1.513	&-1.481	&-1.489	&-1.378	&-1.398	&26&3\\
IC4499       	&-1.53	&-1.573	&-1.584	&-1.641	&-1.628	&-1.707	&-1.669	&-1.670	&-1.635	&23&9\\ 
M3	        &-1.50	&-1.486	&-1.500	&-1.532	&-1.527	&-1.526	&-1.534	&-1.516	&-1.493	&52&10\\
NGC\,6934	&-1.47	&-1.386	&-1.417	&-1.498	&-1.548	&-1.443	&-1.443	&-1.459	&-1.459	&25&2\\
NGC\,6229	&-1.47	&-1.273	&-1.267	&-1.395	&-1.393	&-1.390	&-1.390	&-1.322	&-1.322	&18&2\\
M5	        &-1.29	&-1.285	&-1.304	&-1.364	&-1.383	&-1.380	&-1.395	&-1.356	&-1.363 &35&13\\
M14	        &-1.28	&-1.337	&-1.283	&-1.486	&-1.467	&-1.343	&-1.305	&-1.374	&-1.330 &16&16\\
NGC\,1851	&-1.18	&-1.310	&-1.285	&-1.242	&-1.233	&-1.345	&-1.383	&-1.307	&-1.309	&6&6\\
\hline
\multicolumn{12}{l}{The superscript $^{*}$ denotes photometric metallicities calculated  according to Eqs.~\ref{fe/h.v.eq} and \ref{fe/h.i.eq}.} 
\end{tabular}
\end{center}    
\end{table*}

The  Bailey diagrams in the $V$ and the $I$ bands,  defined by the final sample of stars of the 10 GCs are shown  in Fig~\ref{Bailey.fig}. Different symbols denote stars in different clusters. The  Oo\,I- and Oo\,II-type non-Blazhko RRab stars in the Galactic bulge, as classified by \cite{prud19} using  OGLE-IV data  \citep{sos17}, are also plotted by small dots in the right-hand panel of Fig~\ref{Bailey.fig}.

The classification of the cluster variables to Oo\,I and Oo\,II types is based on the location of the  two Oo types in the $I$-band Bailey diagram  according to the study of \cite{prud19}.
 The division between the two Oo types of the OGLE IV data is defined as:
\begin{equation}\label{ifit.eq}
A(I)= -9.656 P^3 + 11.374 P^2 - 6.327 P +2.398.
\end{equation}

Eq.~\ref{ifit.eq} has been determined by estimating the position of the separating curve at 12 period values between 0.45 d and 0.72 d and fitting them with a third-order polynomial.
\cite{prud19} gave second- and third-order coefficients of the Oo\,I locus defined by the Galactic bulge data, and following \cite{mic08}, accepted a 0.045 d uniform period offset of the Oo\,II sample. However, they used an inverse-form relation and logarithmic period scale [i.e., $\log P(A)$]. The transformation of these coefficients to an $A(P)$ relation yields larger uncertainties than deriving Eq.~\ref{ifit.eq} from their results directly. 

The amplitudes of RRL stars in different photometric bands are strongly correlated, making it possible to transform Eq.~\ref{ifit.eq}  to the $V$ band. The mean value of the $A_V/A_I$ amplitude ratio for the 293 cluster variables involved in this study is 1.525 with a scatter of 0.077. Multiplying Eq.~\ref{ifit.eq} by 1.525 we obtain the following $V$-band division of the Oo groups:
\begin{equation}\label{vfit.eq}
A(V)= -14.726 P^3 + 17.345 P^2 - 9.649 P +3.658.
\end{equation}

The classification of variables lying close to the division of the two Oo types are somewhat contradictory in some cases. The Oo classification of these stars relies primarily on their $V$-band position on the Bailey diagram because, in general, the $V$ light curves are more accurate than the $I$ band ones. The Oo\,I locus of the sample stars of a given cluster is also considered in otherwise questionable cases, as the Oo\,I ridge of the more metal-rich clusters is somewhat below of the Oo\,I ridge of the less metal-rich ones. The position of V26 in M3  is exactly on the separation line both in the $V$- and the $I$-band Bailey diagrams. Its classification as Oo\,II is taken from \cite{jur17}, which was based both on the period/amplitude relation and on the mean magnitude of the variables. 

Mean and median values of the photometric metallicities in the two photometric bands are derived for the Oo\,I- and Oo\,II-type variables separately for the 10 clusters involved in the analysis. The results are summarised in Table~\ref{ered2.tab}.

{\begin{figure*}
\centering
\includegraphics[width=17.2cm]{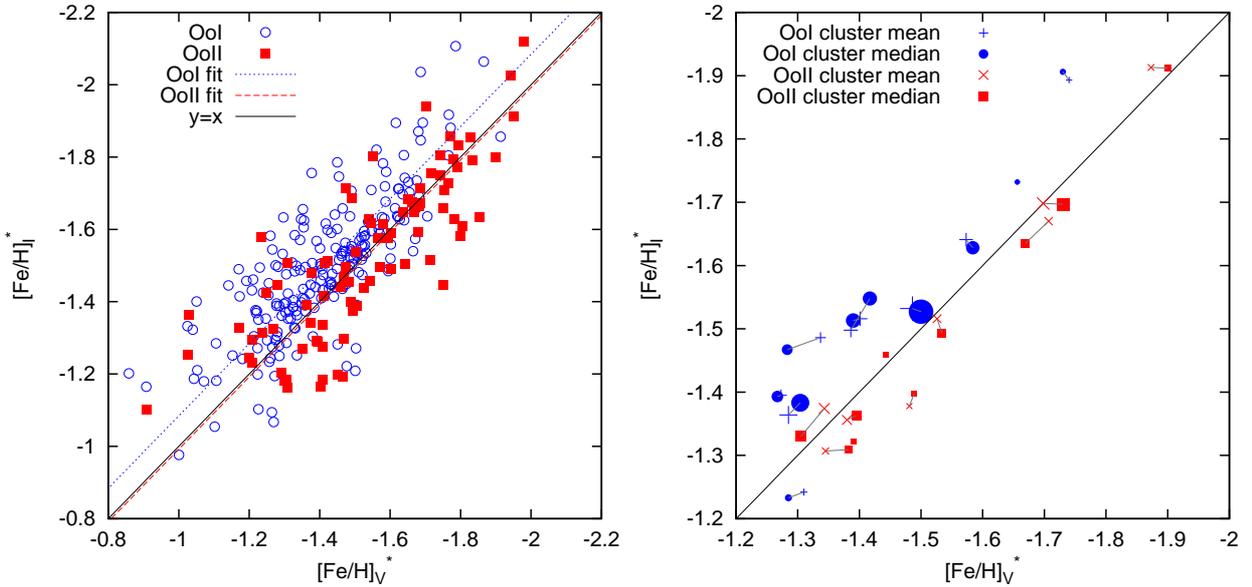}
\caption{ The $V$- and $I$-band photometric metallicities of the total sample of 293 stars (left-side panel), and of the cluster median and mean values (right-side panel) are compared. The fits shown in the left-hand panel surmise constant offsets between the data. The Oo-type\,I and II stars are denoted by different symbols and the symbol sizes are scaled according to the sample size in the right-hand panel. The mean and median [Fe/H] values of the same cluster are connected in this plot. }
\label{all.fig} 
\end{figure*}
\begin{figure}
\centering
\includegraphics[width=9.6cm]{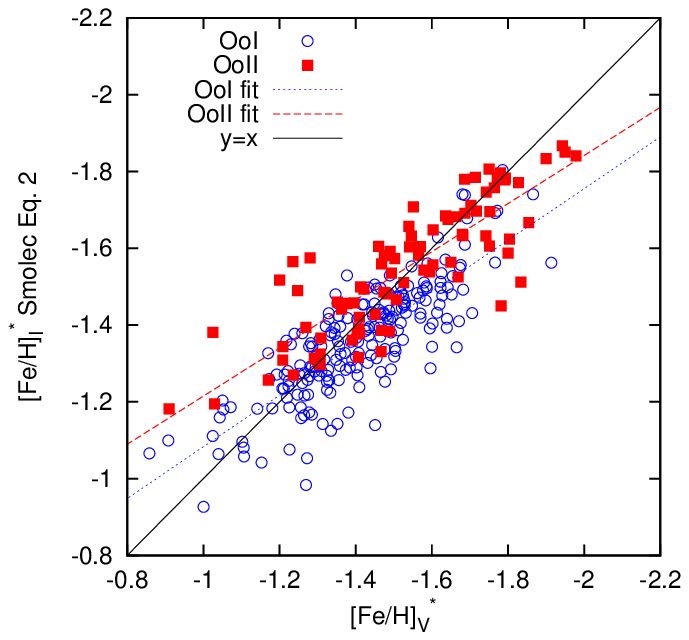}
\caption{ Comparison of photometric metallicities obtained using eq. 2 of
\citet{s05} with
the V-band results. Note the much more significant differences between the $V$- and $I$-band [Fe/H] values of both Oo types comapered to the results shown in Fig.~\ref{all.fig}. The fitted lines show the $\mathrm{[Fe/H]_I=a*[Fe/H]_V+c}$ relations. } 
\label{smo2.fig} 
\end{figure}}

\section{Metallicities of the Oo\,I, Oo\,II groups}

\subsection {Comparison of the $V$- and $I$-band results}\label{OoI-II}

The $V$- and $I$-band photometric metallicities for all the 293 stars and for the cluster mean/median values  are plotted in the left- and the right-hand panels of Fig.~\ref{all.fig}, respectively. The results for Oo\,I and Oo\,II stars are shown with different symbols in these plots.

Systematic differences between the $V$- and $I$-band results of the Oo-type I sample are evident in both plots of  Fig.~\ref{all.fig}.  The results both for individual stars, and for the cluster mean/median values indicate that the $I$-band metallicities are lower than those derived from the $V$-band data for Oo-type I stars, similarly as it was detected for the complete data set. However, the Oo II sample does not show a significant difference between the results in the two photometric bands. Looking at the results of the cluster mean/median values (right-hand panel of Fig.~\ref{all.fig}) the Oo\,II sample shows a marginal, but opposite direction offset.

Fitting the offsets  between the results of the two photometric bands gives $-0.084\pm0.008$ and $0.008\pm0.015$ vales for the Oo\,I and Oo\,II total samples of stars. The sizes of these offsets are also shown by dotted blue and dashed red lines in  Fig.~\ref{all.fig}. The sample-size  weighted mean value of these two offsets is identical with the $-0.057$ offset of the total sample shown in the left-hand panel of Fig.~\ref{meanvas1.fig}. 

Therefore, the explanation of the differences between the  $V$- and $I$-band photometric metallicity values obtained for the global sample of stars is that the formulae do not give identical results when they are applied for the different Oo-type RRL stars in the sense that:
\begin{align}
\mathrm{[Fe/H]}_I &<\mathrm{[Fe/H]}_V & \,\,\, \mathrm{Oo\,I}\\
\mathrm{[Fe/H]}_I &\gtrsim \mathrm{[Fe/H]}_V & \,\,\, \mathrm{Oo\,II}
\end{align}

 In order to justify our decision not using eq.~2 of \cite{s05} to calculate the $I$-band [Fe/H] (see Sect.~\ref{intro}), in Fig. 6 we compare the [Fe/H] values obtained with this formula to the
V-band results, similarly to the left-hand panel in Fig. 6. which in contrast shows the
results for eq. 3 of \cite{s05}.  It can be seen that the differences between the $V$- and $I$-band [Fe/H] values are strongly [Fe/H] dependent in this case. This is in high contrast with the results obtained according to eq.~3 of \cite{s05}, which indicate only an offset between the results and only for the OoI sample.

\subsection {Comparison of the results for Oo\,I and Oo\,II samples}\label{comp2}

We can also revise whether there is any systematic difference between the metallicities of the two Oo types in the same cluster.
This is investigated separately for the $V$- and the $I$-band formulae. Supposing that the globular clusters of our sample are mono-metallic,  no difference between the metallicities of Oo\,I and Oo\,II stars should be detected in a given cluster. Consequently, any systematic difference between their metallicities  has to arise because of biases of the method. 

\begin{figure*}
\centering
\includegraphics[width=16.cm]{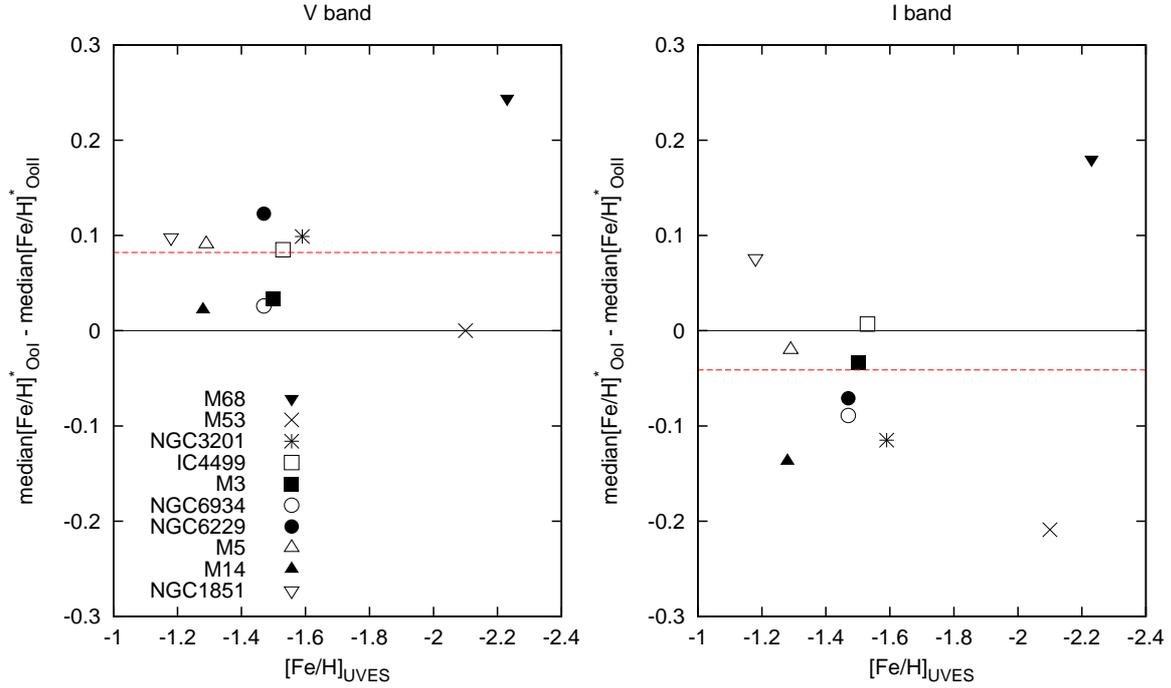}
\caption{Differences between the median values of the [Fe/H]${^*}_{\mathrm{phot}}$ values of Oo\,I and Oo\,II stars versus the UVES metallicities of the clusters are shown for the $V$ and  $I$ bands. 
 The black solid and the red dashed lines indicate the zero difference and the mean velue of the observed median differences in the plots.}
\label{oodifff.fig} 
\end{figure*}

The difference between the median values of the Oo\,I and Oo\,II populations of each cluster is shown in  Fig.~\ref{oodifff.fig}, against the cluster catalogue metallicity.
The left- and right-hand panels illustrate the $V$- and $I$-band results, respectively.

There is a clear tendency for the $V$-band formula to give $\sim 0.1$\,dex lower estimates for the Oo\,II samples of the clusters than for the Oo\,I groups. 

 Although the result for the $I$ band is not as definite as for the $V$ band, the data indicate a small but opposite sign offset between the results for Oo-type I and type II stars. However, there are two clusters, M68 and NGC 1851, with opposite sign results.

Concerning the $V$ band results we conclude that on the average :
\begin{equation}
\mathrm{[Fe/H]_{Oo\,I}} > \mathrm{[Fe/H]_{Oo\,II}}   \hskip 3cm V  \,\,\,\mathrm{band}.\\
\end{equation}
No universal conclusion on the $I$-band results can be drawn, however the majority of the clusters indicate that:
\begin{equation}
\mathrm{[Fe/H]_{Oo\,I}} \lesssim \mathrm{[Fe/H]_{Oo\,II}}  \hskip 3cm I  \,\,\,\mathrm{band}.\\
\end{equation}

The mean differences between the photometric metallicities of the Oo\,I and Oo\,II samples of the the 10 clusters are $+0.08$ and $-0.04$ dex  for the $V$ and the $I$ bands, respectively.

\section{Conclusion/summary}
The photometric metallicity of large samples of RRab stars are utilised to draw conclusions on different populations/groups of stars in several papers \citep[e.g.][]{piet15,piet20,skow16}, however the inherent systematic biases of the method have never been investigated systematically.
Therefore, we performed a thorough check of the compatibility and accuracy of the results of the  $V$- and $I$-band photometric metallicity formulae of RRab stars, defined in  \cite{jk96} and \cite{s05}, using mono-metallic globular cluster data. 

 In order to check the compatibility of the results for Oo-type I and type II variables, clusters hosting both Oo types of RRL stars were selected. As data of mono-metallic clusters with no indication of [Fe/H] spread among the cluster members are involved in the analysis, metallicity differences cannot be the explanation for the occurrence of both Oo types in these clusters. Following the idea of \cite{ldz90} that RRL stars in Oo-type I clusters are zero-age horizontal branch objects, while those in Oo-type II clusters are more evolved, \cite{cs99} also concluded that the Oosterhoff dichotomy is due to evolution. However, other explanations for the structure and luminosity dispersion on the horizontal branch in GCs, which influences the Oo type of the cluster, have also been suggested, e.g., variations in the Helium content \citep{mar18} or multiple stellar populations \citep{jkl99}.

Comparing the results obtained using the $V$- and $I$-band metallicity formulae, we have found that, in general, the $V$-band photometric [Fe/H] is about 0.05 dex more metal poor than the $I$-band results.
The reason for this offset is that the sample is dominated by Oo\,I stars, and the formulae yield results with the opposite sign differences for stars belonging to the two Oo types in the two photometric bands.  Most probably, these systematic differences  are due to the lack of an amplitude term in the $V$ band and a somewhat large value of the coefficient of the amplitude term in the $I$ band.

 We have also checked whether the calibrating samples collected by  \cite{jk96} and \cite{s05} to define the metallicity formulae have comprised both Oo types of variables in significant numbers. The Oo types are determined according to Eq.~\ref{ifit.eq} and Eq.~\ref{vfit.eq}. The percentages of Oo\,II variables are found to be 28 and 35 for the $V$- and $I$-band calibrating samples, respectively. As Oo\,II stars comprise 29 per cent of the sample analysed in this work, no significant difference between the Oo\,I/Oo\,II number ratio is found. Surprisingly enough, no difference between the accuracy of the [Fe/H] formula applied for the Oo\,I and Oo\,II parts of the $V$-band calibrating sample is detected. The mean differences between the observed and calculated [Fe/H] values are 0.00 dex s.dev. 0.13 (58 stars), and -0.01 dex s.dev. 0.14 (23 stars) for the Oo\,I and Oo\,II sub-samples, respectively. This is in contrast with the results shown in Sect.~\ref{comp2}. As there is no $I$ band light curve publicly available for the $\omega$ Cen variables involved in the $I$-band calibration sample of \cite{s05}, this comparison cannot be performed for the $I$ band calibrating data.

The reason of the discrepant results for the cluster and the field samples is, most probably, that in the field sample the Oo\,II stars are more metal poor than the Oo\,I RRLs, but in the cluster data we compared the results for samples supposed to be identical in metallicity. This result also warns that the photometric metallicity values might be biased not only by the Oo types of the variables but this bias might also depend on the physical origin of the Oo behaviour.

A natural conclusion formed based on these results is that new photometric formulae are needed. However, no previous attempt succeeded in improving the $V$- and $I$-band formulae  \citep[e.g.][]{k05}, mostly because of the lack of new, homogeneous spectroscopic data on large samples of RRab stars. A different way for the derivation of such formulae is through the use of globular cluster data, however, the uncertainty of the metallicities of globular clusters is still of the order of some tenths of a dex for some of the clusters rich in RRL stars (e.g. M68: $-2.16$ \citealt{lee05};  $-2.42$ \citealt{sch15}, NGC1851: $-1.11$ \citealt{kov19}; $-1.33$ \citealt{mar14}). Moreover, metallicity-complex globular clusters also exist \citep[for a review read e.g.][]{gav16}, which are not suitable for the calibration of the formulae.

As a by-product of our study we can strengthen one of the results published by \cite{prud19}, namely that the  Oo\,II sample of RRab stars in the Galactic bulge is about 0.1 dex more metal poor than the  Oo\,I group.
As the [Fe/H]$_{\mathrm{phot}}$ derived from $I$-band data of Oo\,II stars has been found to tend to be less metal poor than of the Oo\,I stars in mono-metallic systems, this systematic bias of the method  is of the opposite direction that was detected by \cite{prud19} in the  bulge. Based on this, the metallicities of the Oo\,II sample might be even more metal-poor than as derived by \cite{prud19}.  Because of its large extent and the extreme extinction towards its direction we have no information on the structure of the horizontal branch in the Galactic bulge. Therefore, the most probable explanation of the detected differences between the mean [Fe/H] values of the Oo-type I and Oo-type II populations is that the latter is strongly contaminated by halo interlopers.

In order to prevent inconsistencies between metallicities derived for Oo\,I and  Oo\,II RR Lyrae, we recommend all future studies aiming to establish empirical relations like these to incorporate comparative analysis between Oo\,I and Oo\,II variables. These in turn should be carried out using independent test samples. As demonstrated here, it is straightforward to utilize globular clusters containing variables belonging to both Oosterhoff groups for this purpose.

\section{Acknowledgements}
The anonymous referee is thanked for the helpful comments, and constructive remarks on the manuscript. 
The support of OTKA grants NN-129075 and K-129249 is acknowledged. The research leading to these results has received funding from
the European Research Council (ERC) under the European Unions Horizon 2020 research and innovation programme (grant agreement No.695099)

\section{Data availability Statement}
 The data underlying this article were accessed from the papers referred in Table 1. The derived data generated in this research will be shared on reasonable request to the corresponding author.

{}

\begin{thebibliography}{99}
\bibitem[\protect\citeauthoryear{Arellano Ferro, Bramich \& Giridhar}{2017}]{fer17} Arellano Ferro A., Bramich D. M.,  Giridhar S.,  2017, RMxAA, 53, 121
\bibitem[\protect\citeauthoryear{Arellano Ferro et al.}{2012}]{fer12} Arellano Ferro A., Bramich D. M., Figuera Jaimes R., Giridhar S., Kuppuswamy K., 2012, MNRAS, 420, 1333 
\bibitem[\protect\citeauthoryear{Arellano Ferro et al.}{2014}]{fer14} Arellano Ferro A., Ahumada J. A., Calder\'on J. H., Kains N., 2014, RMxAA, 50, 307
\bibitem[\protect\citeauthoryear{Arellano Ferro et al.}{2015}]{fer15} Arellano Ferro A., Mancera Pi\~na P. E., Bramich D. M.,  Giridhar S.,  Ahumada J. A., Kains N., Kuppuswamy K., 2015, MNRAS, 452, 727 
\bibitem[\protect\citeauthoryear{Ferro et al.}{2016}]{fer16} Arellano Ferro A., Luna A., Bramich D. M., Giridhar S., Ahumada J. A., Muneer S., 2016, Ap\&SS, 361, 175 
\bibitem[\protect\citeauthoryear{Arp}{1955}]{arp55} Arp H. C., 1955, AJ, 60, 317
\bibitem[\protect\citeauthoryear{Bono, Caputo \& Stellingwerf}{1994}]{bon94} Bono G., Caputo F., Stellingwerf R. F., 1994, APJ, 423, 294
\bibitem[\protect\citeauthoryear{Borissova, Catelan \& Valchev}{2001}]{bor01} Borissova J., Catelan M. Valchev T., 2001, MNRAS, 324, 77
\bibitem[\protect\citeauthoryear{Carretta et al.}{2009}]{carretta} Carretta E., Bragaglia A., Gratton R. G., D'Orazi V., Lucatello S., 2009, A\&A, 508, 695
\bibitem[\protect\citeauthoryear{Castellani, Caputo \& Catellani}{2003}]{cast03} Castellani M., Caputo F., Castellani V., 2003, A\&A, 410, 871
\bibitem[\protect\citeauthoryear{Catelan}{2009}]{cat09} Catelan M., 2009, Ap\&SS, 320, 261    
\bibitem[\protect\citeauthoryear{Clement \& Shelton}{1999}]{cs99}      Clement. C. M., Shelton I., 1999,  ApJ, 515, L85
\bibitem[\protect\citeauthoryear{Contreras Pe\~na et al.}{2018}]{cont18} Contreras Pe\~na C., Catelan M., Grundahl F., Stephens A. W., Smith H. A., 2018, AJ, 155, 116 
\bibitem[\protect\citeauthoryear{D\'ek\'any \& Kov\'acs}{2009}]{dek09} D\'ek\'any I., Kov\'acs G., 2009,  A\&A, 508, 803
\bibitem[\protect\citeauthoryear{Gavagnin, Mapelli \& Lake}{2016}]{gav16} Gavagnin E.,  Mapelli M.,  Lake G., 2016, MNRAS 461, 1276
\bibitem[\protect\citeauthoryear{Hajdu et al.}{2018}]{haj18} Hajdu G., D\'ek\'any I., Catelan M., Grebel K.E., Jurcsik J., 2018,  ApJ, 857, 55
\bibitem[\protect\citeauthoryear{Harris}{1996}]{harris} Harris W. E., 1996 (2010 edition), AJ, 112, 1487
\bibitem[\protect\citeauthoryear{Jacyszyn-Dobrzeniecka et al.}{2020}]{jac20} Jacyszyn-Dobrzeniecka A.M. et al., 2020, ApJ, 889, 13
\bibitem[\protect\citeauthoryear{Jang, Kim \& Lee}{2019}]{jkl99} Jang S., Kim, Jenny J., Lee Y-W.,2019, ApJ, 886, 116
\bibitem[\protect\citeauthoryear{Johnson et al.}{2015}]{john15} Johnson C. I., Rich M.R., Pilachowski C.A., Caldwell N., Mateo M., Bailey J.I., Crane J.D., 2015, AJ, 150, 63
\bibitem[\protect\citeauthoryear{Johnson et al.}{2017}]{john17} Johnson C. I., Caldwell N., Rich M.R., Walker M.G., 2017, AJ, 154, 155
\bibitem[\protect\citeauthoryear{Jurcsik}{1995}]{phot} Jurcsik J., 1995, Acta Astron., 45, 653
\bibitem[\protect\citeauthoryear{Jurcsik \& Kov\'acs}{1996}]{jk96} Jurcsik J.,  Kov\'acs G., 1996, A\&A, 312, 111
\bibitem[\protect\citeauthoryear{Jurcsik et al.}{2003}]{j03} Jurcsik, J., Benk\H o, J., Bakos, G., Szeidl, B., Szeb\'o, R. 2003, ApJ, 597, L49
\bibitem[\protect\citeauthoryear{Jurcsik et al.}{2017}]{jur17} Jurcsik J. et al., 2017, MNRAS, 468, 1317 
\bibitem[\protect\citeauthoryear{Kains et al.}{2015}]{kai15}  Kains N. et al., 2015, A\&A, 578, 23 
\bibitem[\protect\citeauthoryear{Kaluzny et al.}{2000}]{kal00}  Kaluzny J., Olech A., Thompson I., Pych W., Krzeminski W., Schwarzenberg-Czerny A., 2000, A\&AS, 143, 215 
\bibitem[\protect\citeauthoryear{Kaluzny et al.}{2001}]{kal01}  Kaluzny J., Olech A., Stanek K. Z., 2001, AJ, 121, 1533 
\bibitem[\protect\citeauthoryear{Kov\'acs}{2005}]{k05} Kov\'acs G., 2005, A\&A, 438, 227
\bibitem[\protect\citeauthoryear{Kovalev et al.}{2019}]{kov19} Kovalev M.,  Bergemann M.,  Ting Y-S.,  Rix H-W., 2019, A\&A,  628, A54 
\bibitem[\protect\citeauthoryear{Landolt}{1983}]{landolt83} Landolt A. U., 1983, AJ, 88, 439
\bibitem[\protect\citeauthoryear{Landolt}{1992}]{landolt92} Landolt A. U., 1992, AJ, 104, 340
\bibitem[\protect\citeauthoryear{Layden \& Sarajedini}{2003}]{lay03} Layden A.C., Sarajedini A., 2003, AJ, 125, 208
\bibitem[\protect\citeauthoryear{Lee, Demarque \& Zinn}{1990}]{ldz90} Lee Y.-W., Demarque P., Zinn R., 1990, ApJ, 350, 155
\bibitem[\protect\citeauthoryear{Lee, Carney \& Habgood}{2005}]{lee05} Lee J-W.,  Carney B.W.,  Habgood M.J., 2005, AJ, 129, 251L
\bibitem[\protect\citeauthoryear{Marconi et al.}{2018}]{mar18} Marconi M., Bono G., Pietrinferni A., Braga V.F.,  Castellani M., Stellingwerf R.F.  2018, ApJ, 864, L13
\bibitem[\protect\citeauthoryear{Marino et al.}{2014}]{mar14} Marino A.F. et al., 2014, MNRAS, 442, 3044
\bibitem[\protect\citeauthoryear{Magurno et al.}{2018}]{mag18} Magurno D., et al., 2018, ApJ, 864, 57
\bibitem[\protect\citeauthoryear{Miceli et al.}{2008}]{mic08} Miceli A., 2008, ApJ, 678, 865
\bibitem[\protect\citeauthoryear{Morgan, Wahl \& Wieckhorst}{2007}]{mor07} Morgan S.M.,  Wahl, J.N.,  Wieckhorst R.M., 2007, MNRAS, 374, 1421
\bibitem[\protect\citeauthoryear{Mullen et al.}{2021}]{mul21} Mullen J.P. et al., 2021, arXiv:2103.09372
\bibitem[\protect\citeauthoryear{Nemec}{2004}]{nem04} Nemec J. M.,  2004, AJ, 127, 2185
\bibitem[Nemec et al.(2013)]{nemec} Nemec, J. M., Cohen, J., Ripepi, V., Derekas, A., Moskalik, P., Sesar, B., Chadid, M., Bruntt, H., 2013, ApJ, 773, 181
\bibitem[\protect\citeauthoryear{Oosterhoff}{1939}]{oo39} Oosterhoff P. T., 1939, Obs, 62, 104
\bibitem[\protect\citeauthoryear{Pietrukowicz et al.}{2015}]{piet15} Pietrukowicz P. et el., 2015, ApJ, 811, 113
\bibitem[\protect\citeauthoryear{Pietrukowicz et al.}{2020}]{piet20} Pietrukowicz P. et el., 2020, Acta Astron., 70, 121
\bibitem[\protect\citeauthoryear{Prudil et al.}{2019}]{prud19} Prudil Z., D\'ek\'any I., Catelan M., Smolec R., Grebel E.K., Skarka M., 2019, MNRAS, 484, 4833
\bibitem[\protect\citeauthoryear{Reid}{1996}]{reid} Reid N., 1996, MNRAS, 278, 367
\bibitem[\protect\citeauthoryear{Schaeuble}{2015}]{sch15} Schaeuble M., Preston G., Sneden C., Thompson I. B., Shectman S. A., Burley G. S., 2015, AJ, 149, 204
\bibitem[\protect\citeauthoryear{Skowron et al.}{2016}]{skow16} Skowron D.M. et al., 2016, Acta Astron., 66, 269
\bibitem[\protect\citeauthoryear{Smolec}{2005}]{s05} Smolec R., 2005, Acta Astron., 55, 59
\bibitem[\protect\citeauthoryear{Soszy\'nski et al.}{2017}]{sos17} Soszy\'nski I. et al., 2017, Acta Astron., 67, 297 
\bibitem[\protect\citeauthoryear{Stetson}{2000}]{stet00} Stetson P.B., 2000, PASP, 112, 925
\bibitem[\protect\citeauthoryear{Torrealba et al.}{2015}]{tor15} Torrealba G. et al., 2015, MNRAS, 446, 2251
\bibitem[\protect\citeauthoryear{Walker}{1994}]{wal94} Walker A. R., 1994, AJ, 108, 555
\bibitem[\protect\citeauthoryear{Walker \& Nemec}{1996}]{wal96} Walker A. R., Nemec J. M., 1996, AJ, 112, 2026
\bibitem[\protect\citeauthoryear{Walker}{1998}]{wal98} Walker A. R., 1998, AJ, 116, 220
\bibitem[\protect\citeauthoryear{Yepez et al.}{2018}]{ye18} Yepez M. A., Arellano Ferro A., Muneer S., Giridhar S., 2018, RMxAA, 54, 15 
\bibitem[\protect\citeauthoryear{Yong et al.}{2014}]{yong14} Yong D., 2014, MNRAS, 441, 3396
\end{thebibliography}
\end{document}